\documentclass[aps,prd,twocolumn,groupedaddress,nofootinbib,floatfix]{revtex4}

\usepackage{graphicx}
\usepackage{bbm}
\usepackage{amsmath}
\usepackage{amssymb}
\usepackage{comment}
\usepackage{cancel}
\DeclareGraphicsExtensions{.eps, .eps, .jpg, .png}

\def\by{{\bar{y}}}

\def\bk{{\bf k}}
\def\bx{{\bf x}}
\def\bp{{\bf p}}

\def\by{{\bf y}}

\begin{document}
%\tighten
%%\draft
%\renewcommand{\topfraction}{0.8}

%\preprint{}

% Real Title of paper
%\title{Next-generation test of cosmic inflation}

% Fake title for internal use

\title{The spectrum of tensor perturbations in warm inflation}


%\author{}

\author{Yue Qiu}

\email[]{yqiu@umass.edu}
\author{Lorenzo Sorbo}

\email[]{sorbo@physics.umass.edu}

%\thanks{}
\affiliation{Amherst Center for Fundamental Interactions, Department of Physics, University of Massachusetts, Amherst, MA 01003, U.S.A.}
%\thanks{}


\date{\today}

\begin{abstract}
We compute the spectrum of tensor perturbations in warm inflation. We find that the spectrum, besides the standard component $\propto {H^2}/{M_P^2}$ associated to the amplification of the tensor vacuum fluctuations, acquires a component $\propto {\ell_{\rm mfp}\,T^5}/{M_P^4}$, where $\ell_{\rm mfp}$ and $T$ are respectively the mean free path and the temperature of the thermal degrees of freedom. The new contribution is due to the direct production of gravitational waves by the thermal bath, and can exceed the standard one in a viable region of parameter space. This contribution is dominated by thermal fluctuations at scales longer than $\ell_{\rm mfp}$. 
\end{abstract}

%\keywords{}

\maketitle


%%%%%%%%%%%%%%%%%%%%%%%
\section{Introduction}%
\label{sec:intro}%%%%%%
%%%%%%%%%%%%%%%%%%%%%%%

In models of warm inflation~\cite{Berera:1995wh,Berera:1995ie}, the inflaton interacts with a thermal bath of relativistic particles with a slowly evolving temperature $T$. In order to prevent the temperature from redshifting away, the thermal bath must be continuously replenished by some interaction with the inflaton - the form of interactions being model dependent.   The spectrum of metric scalar perturbations in warm inflation has been studied in several works, see e.g.~\cite{Berera:1995wh,Berera:1995ie,Hall:2003zp,Graham:2009bf}, and its expression depends on the specific form of the interaction between the thermal bath and the inflaton. The tensor perturbations will see, as usual, their vacuum fluctuations amplified by the accelerated expansion, which will lead to a contribution to the their power spectrum with amplitude ${\cal P}^t_{\rm vac}=\frac{2}{\pi^2}\frac{H^2}{M_P^2}$. The thermal bath will provide an additional source of tensors. In this work we compute this contribution. 

Since the interaction of gravitational waves with the thermal bath depends only on the properties of the latter, our results will not depend on the specifics of the inflaton sector. It will however depend on the strength of the interactions that maintain the thermal bath in equilibrium.

Besides the Hubble parameter $H$ and the temperature $T$, a relevant scale for our system will be given by the mean free path $\ell_{\rm mfp}$ of the particles in the thermal bath. For thermal inflation to be at work, the hierarchy $T\gtrsim \ell_{\rm mfp}^{-1}\gg H$ must be realized. The first inequality derives from the fact that one cannot define a mean free path shorter than the thermal wavelength, the second is equivalent to the requirement of thermal equilibrium in an expanding Universe. In Section~\ref{sec:short} we will compute the contribution to the tensor spectrum from modes at length scales much shorter than $\ell_{\rm mfp}$, whereas in Section~\ref{sec:long} we will compute the contributions from larger scales, that will give the dominant effect.

%%%%%%%%%%%%%%%%%%%%%%%%%%%%%%%%%%%%%%
\section{The sourced tensor spectrum}%
%%%%%%%%%%%%%%%%%%%%%%%%%%%%%%%%%%%%%%

We work in conformal time, and consider only transverse-traceless perturbations $h_{ij}(\bx,\,\tau)$ around a flat Friedmann-Robertson-Walker background $ds^2=a(\tau)^2\left[-d\tau^2+\left(\delta_{ij}+h_{ij}\right)dx^idx^j\right]$. We will approximate the inflating Universe with a de Sitter space, $a(\tau)=-1/(H\tau)$. Then, in the presence of a stress-energy tensor $T_{ab}(\bx,\,\tau)$, that we assume to be generated by a bath of relativistic particles,  the tensor fluctuations satisfy the equation
\begin{align}\label{eq:eq1}
h_{ij}''(\bx,\,\tau)+2\frac{a'}{a}\,&h_{ij}'(\bx,\,\tau)-\Delta h_{ij}(\bx,\,\tau)\nonumber\\
&=\frac{2}{M_P^2}\Pi_{ij}{}^{ab}(\partial_\bx)\,T_{ab}(\bx,\,\tau)\,,
\end{align}
where $\Pi_{ij}{}^{ab}(\partial_\bx)=\Pi_i^a(\partial_\bx)\,\Pi_j^b(\partial_\bx)-\frac12 \Pi_{ij}(\partial_\bx)\,\Pi^{ab}(\partial_\bx)$ is the projector on the transverse-traceless modes, with $\Pi_{ij}(\partial_\bx)=\delta_{ij}-\partial_i\partial_j/\Delta$, while a prime denotes a derivative with respect to the conformal time $\tau$. The stress-energy tensor is defined in such a way that $T_{ab}\sim \partial_a\phi\,\partial_b\phi+\ldots$ for a scalar field whose kinetic term is normalized as $\int d\tau\,d^3\bx\,\frac{a^2}{2}\phi'{}^2$. A transformation to canonically normalized fields brings $T_{ab}\rightarrow \frac{1}{a^2}T_{ab}^{(c)}$, where the index ${}^{(c)}$ refers to comoving quantities. Note that eq.~(\ref{eq:eq1}) does not assume thermalization of the gravitational waves. This possibility has been considered in~\cite{Ferreira:2017lnd}   where it was shown that such a situation cannot be achieved consistently in warm inflation.

After taking the Fourier transform of eq.~(\ref{eq:eq1}), and solving it in terms of the Green's function $G_p(\tau,\,\tau')$, we obtain the correlator
\begin{align}\label{eq:main_corr}
&\langle h_{ij}(\bp,\,\tau)h_{ij}(\bp',\,\tau)\rangle_{\rm s}=\frac{4}{M_P^4}\int^\tau \frac{d\tau'}{a(\tau')^2}\int^\tau \frac{d\tau''}{a(\tau'')^2}\nonumber\\
&\quad\times G_p(\tau,\,\tau')\,G_{p'}(\tau,\,\tau'')\,\Pi_{ij}{}^{ab}(-i\bp)\,\Pi_{ij}{}^{cd}(-i\bp')\nonumber\\
&\quad\times\int \frac{d^3\bx\,d^3\bx'}{(2\pi)^{3}}e^{-i\bp\bx-i\bp'\bx'}\langle T{}^{(c)}_{ab}(\bx,\,\tau')\,T{}^{(c)}_{cd}(\bx',\,\tau'')\rangle\,,
\end{align}
where $\langle ...\rangle_{\rm s}$ refers to the component of the correlator sourced by the thermal bath, and where the propagator, in the approximation of exact de Sitter background, reads
\begin{align}
G_p(\tau,\tau')&=\frac{1}{p^3\,\tau'{}^2}\Big[\left(1+p^2\,\tau\,\tau'\right)\sin \left(p\left(\tau-\tau'\right)\right)\nonumber\\
& - \left(p\left(\tau-\tau'\right)\right) \,\cos \left(p\left(\tau-\tau'\right)\right)\Big]\,\Theta\left(\tau-\tau'\right)\,.
\end{align}

In what follows we will consider the tensor spectrum evaluated at the end of inflation, $\tau=-1/H$, at large scales $p\ll H$, so that we will set $\tau=0$ in the propagator.

%%%%%%%%%%%%%%%%%%%%%%%%%%%%%%%%%%%%%%%%%%%%%%%%%%%%
\section{Contribution from short wavelength modes}%%
\label{sec:short}%%%%%%%%%%%%%%%%%%%%%%%%%%%%%%%%%%%
%%%%%%%%%%%%%%%%%%%%%%%%%%%%%%%%%%%%%%%%%%%%%%%%%%%%

Let us start by computing the contribution to the graviton two point function from the stress-energy correlators when both comoving distances and (conformal) time differences are much shorter than the comoving mean free path $\ell^{(c)}_{\rm mfp}$. In this regime we can neglect the effects of interactions and treat our theory as that of a free field. 

For definiteness we will assume that our system is given by a conformally coupled, canonically normalized massless scalar field $\varphi$ in thermal equilibrium at comoving temperature $T^{(c)}$. As a consequence, the stress-energy tensor correlator appearing in eq.~(\ref{eq:main_corr}) takes the form
\begin{widetext}
\begin{align}
\langle T{}^{(c)}_{ab}(\bx,\,\tau')T{}^{(c)}_{cd}(\bx',\,\tau'')\rangle=&\partial_{y_1^a}\partial_{y_2^b}\partial_{y_3^c}\partial_{y_4^d}\Big[\langle\varphi(\by_1,\,\tau')\varphi(\by_2,\,\tau')\varphi(\by_3,\,\tau'')\varphi(\by_4,\,\tau'')\rangle\nonumber\\
&-\langle\varphi(\by_1,\,\tau')\varphi(\by_2,\,\tau')\rangle\,\langle\varphi(\by_3,\,\tau'')\varphi(\by_4,\,\tau'')\rangle\Big]\Big|_{\by_1=\by_2=\bx,\,\by_3=\by_4=\bx'}\,,
\end{align}
\end{widetext}
where we ignored the part of stress-energy tensor proportional to $\delta_{ab}$ that is projected out by $\Pi_{ij}{}^{ab}(\partial_\bx)$.

To compute $\langle\varphi(\by_1,\,\tau')\,\varphi(\by_2,\,\tau')\,\varphi(\by_3,\,\tau'')\,\varphi(\by_4,\,\tau'')\rangle$ in a thermal state we Wick-rotate to Euclidean spacetime with periodic imaginary (conformal) time, $\varphi(i\tau+1/T^{(c)})=\varphi(i\tau)$ and we use Wick's theorem to decompose the four-point correlator into products of thermal Green's functions. The thermal Green's function at comoving temperature $T^{(c)}$, in terms of the Euclidean conformal time $\tau_E=i\tau$ reads
\begin{align}
&G_T(x,\,\tau_E)=-T^{(c)}\int \frac{d^3\bk}{(2\pi)^3}\sum_{n=-\infty}^\infty \frac{e^{2\pi i n T^{(c)}\tau_E+i\bk\cdot\bx}}{(2\pi n\,T^{(c)})^2+\bk^2}\nonumber\\
&\qquad=-\frac{T^{(c)}}{4\pi\, x}\frac{\sinh\left(2\pi T^{(c)} x\right)}{\cosh\left(2\pi T^{(c)} x\right)-\cos\left(2\pi T^{(c)} \tau_E\right)}\,,
\end{align}
that, rotating back to real conformal time, turns into~\cite{Eftekharzadeh:2010qp}
\begin{align}
&G_T(x,\,\tau)=-\frac{T^{(c)}}{4\pi\,x}\frac{\sinh\left(2\pi T^{(c)} x\right)}{\cosh\left(2\pi T^{(c)} x\right)-\cosh\left(2\pi T^{(c)} \tau\right)}\,.
\end{align}

Note that in the limit $T^{(c)}\to 0$ we obtain the Minkowskian Green's function for a massless field,
\begin{align}
&G_0(x,\,\tau)=-\frac{1}{4\pi^2}\,\frac{1}{x^2-\tau^2}\,.
\end{align}
To renormalize away the effects of the zero temperature fluctuations of $\varphi$, we will work with the subtracted Green's function
\begin{align}
G^{\rm sub}_T(x,\,\tau)=G_T(x,\,\tau)-G_0(x,\,\tau)\,.
\end{align}

We are now in position to compute
\begin{align}
&\langle T{}^{(c)}_{ab}(\bx,\,\tau)T{}^{(c)}_{cd}({\bf 0},\,0)\rangle=2\,\hat\bx_a\hat\bx_b\hat\bx_c\hat\bx_d\,G_{T,\,xx}^{\rm sub}(x,\,\tau'-\tau'')^2,
\end{align}
where we denote $G_{T,\,xx}^{\rm sub}(x,\,\tau)=\partial_x^2\,G^{\rm sub}_T(x,\,\tau)$, and where a hat denotes a vector with unit length. We thus obtain
\begin{align}\label{eq:hh_noscattering}
&\langle h_{ij}(\bp,\,0)\,h{}^{ij}(\bp',\,0)\rangle_{\rm s,\ short}=\frac{4}{M_P^4}\delta(\bp+\bp')\nonumber\\
&\times\int \frac{d\tau'}{a(\tau')^2}\int \frac{d\tau''}{a(\tau'')^2} G_p(0,\,\tau')\,G_{p'}(0,\,\tau'')\, {\cal I}(p,\,\tau'-\tau''),
\end{align}
where we have defined 
\begin{align}
&{\cal I}(p,\,\Delta\tau)\equiv 2\,\Pi_{ij}{}^{ab}(-i\bp)\,\Pi^{ij}{}^{cd}(-i\bp)\nonumber\\
&\qquad\times\int d^3\bx\,e^{-i\bp\cdot\bx}\,\hat{\bx}_a\,\hat{\bx}_b\,\hat{\bx}_c\,\hat{\bx}_d\,G_{T,\,xx}^{\rm sub}(x,\,\Delta\tau)^2\,.
\end{align}
Since here we are considering only the short-distance modes, the upper limit of integration in $d\bx$ in the integral above is given by $\approx \ell_{\rm mfp}^{(c)}$, but, since $G^{\rm sub}_T(x,\,\tau)\to 0$ for $2\pi T^{(c)}\,x\gtrsim 1$, we can approximate it by infinity assuming $2\pi T^{(c)}\,\ell_{\rm mfp}^{(c)}\gg 1$.

\begin{figure}[h]
\centering
\includegraphics[scale=.6]{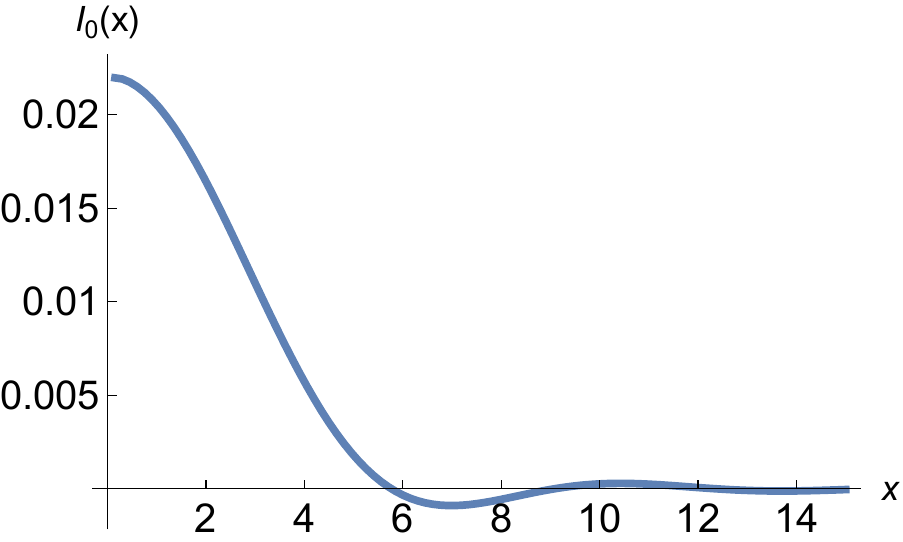} 
\caption{The function ${\cal I}_0(x)$, defined in eq.~(\ref{eq:calI_num}).}
\label{fig:i0}
\end{figure}

Numerical evaluation then gives that for $p\ll 2\pi T^{(c)}$, 
\begin{align}\label{eq:calI_num}
{\cal I}(p,\,\Delta\tau)\simeq {\cal I}_0(p|\Delta\tau|)\,\, T^{(c)}{}^5\,,
\end{align}
where the function ${\cal I}_0(x)$ is plotted in Figure~\ref{fig:i0}. The modes with $p\gtrsim 2\pi T^{(c)}$ are suppressed and irrelevant.

%%%%%%%%%%%%%%%%
%%%%%%%%%%%%%%%%
%%%%%%%%%%%%%%%%
%%%%%%%%%%%%%%%%

The comoving temperature $T^{(c)}$ appearing in eq.~(\ref{eq:calI_num}) is  time-dependent, as it is given by $a\,T$, where the physical temperature $T$ is approximately constant during warm inflation. This raises the question of whether $T^{(c)}$ should be evaluated at time $\tau'$ or at time $\tau''$. The fact that we are considering short distance modes helps us here. In fact, for those modes $|\tau'-\tau''|\lesssim \ell_{\rm mfp}^{(c)}=-\tau' \left(\ell_{\rm mfp}H\right)$ and since thermalization requires $\left(\ell_{\rm mfp}H\right)\ll 1$, we have $|\tau'-\tau''|\ll |\tau'|\simeq |\tau''|$ in our integral. As a consequence, the short wavelength contribution to the graviton correlator will be confined to the region of integration with $\tau'\simeq\tau''$ and it makes no difference whether $T^{(c)}$ is evaluated at $\tau'$ or $\tau''$. To keep things symmetric, we will assume $T^{(c)}=\sqrt{a(\tau')\,a(\tau'')}\,T$ inside the integral.

The condition $|\tau'-\tau''|\lesssim \ell_{\rm mfp}^{(c)}$ also helps to simplify the next step. Since the propagators multiplied by the factor $T^{(c)}{}^5=a(\tau')^{5/2}\,a(\tau'')^{5/2}\,T^5$ give suppressed contribution unless $|p\tau'|\approx |p\tau''|={\cal O}(1)$, we obtain that $p|\tau'-\tau''|\lesssim |p\tau'| \left(\ell_{\rm mfp}H\right)={\cal O}\left(\ell_{\rm mfp}H\right)\ll 1$, so that we can approximate 
\begin{align}
{\cal I}(p,\,\Delta\tau)\approx \left\{
\begin{array}{ll}
{\cal I}_0(0)\, T^{(c)}{}^5\simeq .02\, T^{(c)}{}^5, & p|\Delta\tau|\lesssim \left(\ell_{\rm mfp}H\right)\\
0, &p|\Delta\tau|\gtrsim \left(\ell_{\rm mfp}H\right).
\end{array}
\right. 
\end{align}

We finally find the approximate result
\begin{align}
&\int^\tau \frac{d\tau'}{a(\tau')^2}\int^\tau \frac{d\tau''}{a(\tau'')^2} G_p(\tau,\,\tau')\,G_{p'}(\tau,\,\tau'')\, {\cal I}(p,\,\tau'-\tau'')\nonumber\\
&\qquad\approx \frac{\left(\ell_{\rm mfp}H\right)}{p}\int^\tau \frac{d\tau'}{a(\tau')^4} G_p(\tau,\,\tau')^2\times .02\, T^{(c)}{}^5\nonumber\\
&\qquad= \frac{5\times 10^{-3}}{p^3}\ell_{\rm mfp}\,T^5\,.
\end{align}
Introducing the tensor power spectrum ${\cal P}^t$ through $\langle h_{ij}(\bp,\,\tau)h_{ij}(\bp',\,\tau)\rangle=\frac{2\pi^2}{p^3}{\delta^{(3)}(\bp+\bp')}\,{\cal P}^t(p)$, we finally obtain
\begin{align}\label{eq:Ptshort}
{\cal P}^t_{\rm s,\ short}(p)\approx 10^{-3}\frac{\ell_{\rm mfp}\,T^5}{M_P^4}\,.
\end{align}

We will now consider the contribution from hydrodynamic modes with wavelength larger than the mean free path, and we will find that they give the dominant contribution to the sourced correlator.

%%%%%%%%%%%%%%%%%%%%%%%%%%%%%%%%%%%%%%%%%%%%%%%%
\section{Contribution from hydrodynamic modes}%%
\label{sec:long}%%%%%%%%%%%%%%%%%%%%%%%%%%%%%%%%
%%%%%%%%%%%%%%%%%%%%%%%%%%%%%%%%%%%%%%%%%%%%%%%%

In the hydrodynamic regime (in which either distances or time differences are larger than the mean free path of the particles) we can apply a treatment analogous to that used in~\cite{Ghiglieri:2015nfa,Ghiglieri:2020mhm} for the case of a radiation dominated Universe. We start from the relation~\cite{lifschitz}
\begin{align}
&\langle T{}^{(c)}_{ab}(\bx,\,\tau)\,T{}^{(c)}_{cd}(\bx',\,\tau')\rangle=2\,T{}^{(c)}\,\Bigg[\eta{}^{(c)}\,(\delta_{ac}\,\delta_{bd}+\delta_{ad}\,\delta_{bc})\nonumber\\
&\quad \quad +\left(\zeta{}^{(c)}-\frac23\eta{}^{(c)}\right)\delta_{ab}\,\delta_{cd}\Bigg]\delta(\bx-\bx')\,\delta(\tau-\tau')\,,
\end{align}
where $\eta{}^{(c)}$ and $\zeta{}^{(c)}$ are respectively the comoving shear and the bulk viscosity. Inserting the expression above into eq.~(\ref{eq:main_corr}) we obtain
\begin{align}\label{eq:final_hydro_pt}
{\cal P}^t_{\rm s,\ long}(p)=\frac{24\,p^3}{\pi^2\,M_P^4}\int \frac{d\tau'}{a(\tau')^4} G_p(0,\,\tau')^2\,T^{(c)}(\tau')\,\eta^{(c)}(\tau'),
\end{align}
where we used $\Pi_{ij}{}^{ab}(-i\bp)\,\Pi_{ij}{}^{ab}(-i\bp)=3$.

Eq.~(\ref{eq:final_hydro_pt}) is our main result. To proceed we need to specify the expression of $\eta^{(c)}$, that depends on the details of the interactions within the thermal bath. 

The shear viscosity can take values between two limits. 

A lower bound on $\eta^{(c)}$ is conjectured~\cite{Kovtun:2004de} to be
\begin{align}\label{eq:eta_small}
\eta^{(c)}\ge \frac{s^{(c)}}{4\pi}\,,
\end{align}
where $s^{(c)}=\frac{2\pi^2}{45}g_{*,S}\, T^{(c)}{}^3$ is the comoving entropy density of the thermal gas, with $g_{*,S}$ denoting the effective number of degrees of freedom in entropy. Applying the inequality~(\ref{eq:eta_small}), we obtain
\begin{align}\label{eq:lower_pt1}
{\cal P}^t_{\rm s,\ long}(p)\ge \frac{4}{15\pi M_P^4}\,g_{*,S}\,T{}^4\,p^3\int d\tau'\,G_p(0,\,\tau')^2
\end{align}
where we have assumed that the physical temperature $T=T^{(c)}/a$ is approximately constant. Evaluation of the integral in $d\tau'$ gives 
\begin{align}\label{eq:lower_pt2}
{\cal P}_{\rm s,\ long}^t(p)\ge \frac{2}{45\,M_P^4}g_{*,S}\,T^4\simeq \frac{4}{3\pi^2}\frac{\rho_r}{M_P^4}
\end{align}
where in the last step we have introduced the energy density in the radiation, $\rho_r=\frac{\pi^2}{30}g_*\,T^4$ assuming $g_*\simeq g_{*,S}$.

Since by assumption the radiation must be subdominant with respect to the inflaton energy, $\rho_r\ll 3\,H^2M_P^2$, eq.~(\ref{eq:lower_pt2}) shows that if the inequality~(\ref{eq:eta_small}) is saturated, ${\cal P}^t_{\rm s,\ long}\ll {\cal P}^t_{\rm vac}\equiv\frac{2}{\pi^2}\,\frac{H^2}{M_P^2}$.

An upper bound on ${\cal P}^t_{\rm s,\ long}$ is induced by an upper bound on $\eta^{(c)}$. The shear viscosity is approximately given by
\begin{align}
\eta^{(c)}\approx \ell_{{\rm mfp}}^{(c)}\,T^{(c)}{}^4\,.
\end{align}
Imposing that the mean free path is much shorter than the horizon radius $\ell_{{\rm mfp}}^{(c)}\ll (a\,H)^{-1}$, we obtain the upper bound
\begin{align}
{\cal P}^t_{\rm s,\ long}(p)\ll \frac{p^3}{M_P^4}\int^\tau \frac{d\tau'}{a(\tau')^4} G_p(\tau,\,\tau')^2\,\frac{T^{(c)}(\tau')^5}{a(\tau')\,H}\simeq \frac{T^5}{H\,M_P^4}
\end{align}
that, for relatively large values of the temperature, can exceed $ {\cal P}^t_{\rm vac}$ even in a regime in which the energy density in radiation is subdominant with respect to that in the background, $T\ll \sqrt{H\,M_P}$.

%%%
\subsection{An example}
%%%

To work out a specific example, let us consider a model where the thermal bath is given by a real scalar field $\varphi$ with negligible mass and with self-interaction $V(\varphi)=\frac{\lambda}{4!}\,\varphi^4$.

The shear viscosity for this model, in the $\lambda\ll 1$ limit, was computed in~\cite{Jeon:1994if}, where was found that $\eta^{(c)}\simeq 2860\,T^{(c)}{}^3/\lambda^2$. The mean free path is given by $\left[\ell_{\rm mfp}^{(c)}\right]^{-1}=\sigma^{(c)}\,n^{(c)}$, where for a relativistic boson the comoving number density reads $n^{(c)}=\frac{\zeta(3)}{\pi^2}T^{(c)}{}^3$ and the cross section is $\sigma^{(c)}= \frac{\lambda^2}{32\pi^2\,s_{\rm Man}^{(c)}}\simeq\frac{\lambda^2}{128\pi^2\,T^{(c)}{}^2}$ (using the comoving Mandelstam invariant $s_{\rm Man}^{(c)}\simeq (2T^{(c)})^2$). 

%[If I use the thermally averaged cross section I get $\langle \sigma\rangle_T\simeq 7.4\times 10^{-4}\frac{\lambda^2}{T^2}|\log\lambda|$ where the log comes from the thermal mass $\simeq \lambda T^2$ that provides an infrared cutoff to the collinear divergence, that is essentially the same result as above if I set $\log\lambda=1$.)

Using these formulae we obtain
\begin{align}
\eta^{(c)}\simeq 2860\,T^{(c)}{}^3\times\frac{\zeta(3)}{128\pi^4}\ell_{{\rm mfp}}^{(c)}\,T^{(c)}\,\simeq .2\, \ell_{{\rm mfp}}^{(c)}\,T^{(c)}{}^4
\end{align}
and going back to physical quantities, we finally obtain
\begin{align}\label{eq:Ptexample}
{\cal P}^t_{\rm s\ long}(p)\simeq .3\,\frac{\ell_{{\rm mfp}}\,T^5}{M_P^4}
\end{align}
where thermalization requires the model-dependent quantity $\ell_{{\rm mfp}}\ll 1/H$. Comparison of the amplitude of eq.~(\ref{eq:Ptexample}) with that of eq.~(\ref{eq:Ptshort}) shows that the hydrodynamic modes dominate the sourced component of the tensor spectrum.

\smallskip

If, to fix ideas, we set $\ell_{{\rm mfp}}H\simeq .2$, we see that a tensor spectrum as large as $\sim 10^{-10}$ (that saturates the current observational bounds) can be obtained for temperatures $T\simeq 10^{13}\left(H/{\rm GeV}\right)^{1/5}$GeV, where the condition that the radiation density is subdominant by a factor of at least $5$ with respect to the background inflaton energy $\simeq 3\, H^2\,M_P^2$ allows for a Hubble parameter during inflation as low as $\sim 2\times 10^{12}$~GeV. For such a value of the Hubble parameter one gets ${\cal P}^t_{\rm vac}\simeq 10^{-12}$. For this choice of parameters, therefore, the presence of the thermal bath enhances the tensor spectrum by about $2$ orders of magnitude.

\bigskip

To sum up, we have found that the spectrum of gravitational waves generated during thermal inflation includes a component $\propto \ell_{\rm mfp}\,T^5/M_P^4$, sourced by long wavelength thermal modes, that can dominate over the vacuum component in a viable region of parameter space. Our analysis has been agnostic regarding perturbations in the scalar sector, that depend on the details of the interactions between the thermal bath and the inflaton. For this reason, in particular, we give no expression of the amplitude of the tensor-to-scalar ratio (note however that~\cite{Mirbabayi:2014jqa} discussed how mechanisms sourcing tensor modes will generally source scalar perturbations with higher efficiency). It should also be noted that this mechanism might lead to large amplitude of tensor modes towards the end of inflation (where one expect the effects of temperature to be more important) that might be detectable by gravitational interferometers, as discussed for instance in~\cite{Cook:2011hg}.

\acknowledgements We thank Paul Anderson for useful discussions. This work is partially supported by the US-NSF grant PHY-1820675.

\end{document}